\documentclass[conference]{IEEEtran}
\usepackage{cite}
\newcommand{\CLASSINPUTtoptextmargin}{2.4cm}
\usepackage{amsmath,amssymb,amsfonts}
\usepackage{graphicx}
\usepackage{textcomp}
\usepackage{xcolor}
\usepackage{soul}
\usepackage{halloweenmath}
\usepackage{comment}
\usepackage{algpseudocode}
\usepackage{comment}
\usepackage{svg}
\usepackage{multirow}
\algrenewcommand\algorithmicindent{1.0em}
\usepackage[ngerman]{babel}
\usepackage{stackengine}
\usepackage{scalerel}
\usepackage{lipsum}  
\usepackage{tcolorbox}
\usepackage[linesnumbered,lined,ruled,longend, noend]{algorithm2e}

\def\wdot{\textcolor{white}{.}}
\DeclareFontFamily{U}{mathx}{}
\DeclareFontShape{U}{mathx}{m}{n}{<-> mathx10}{}
\DeclareSymbolFont{mathx}{U}{mathx}{m}{n}   
\DeclareMathAccent{\widehat}{0}{mathx}{"70}
\DeclareMathAccent{\widecheck}{0}{mathx}{"71}
\DeclareMathAccent{\widetilde}{0}{mathx}{"72}

\SetCommentSty{mycommfont}
\SetKwInput{KwInput}{Input}                
\SetKwInput{KwOutput}{Output}              

\usepackage{fancyhdr,lipsum}

\fancypagestyle{mahmood}{%
   \fancyhf{} 
   
   \fancyhead[C]{You can use this material personally. Reprinting or republishing this material for the purpose of advertising or promotion, creating new collective works, reselling or redistributing to servers or lists, or using any copyrighted component in other works must adhere to IEEE policy. The DOI will be supplied as soon as it becomes available.}
}%

\makeatletter
\let\ps@IEEEtitlepagestyle\ps@mahmood
\makeatother

\begin{document}
\def\abstractname{Abstract}
\def\refname{References}
\def\figurename{Figure}

\title{ORIENT: A Priority-Aware Energy-Efficient Approach for Latency-Sensitive Applications in 6G}


\author{
    \IEEEauthorblockN{
        Masoud Shokrnezhad\textsuperscript{1}, and Tarik Taleb\textsuperscript{1, 2} \\
    }
    \IEEEauthorblockA{
        \textsuperscript{1} \textit{Oulu University, Oulu, Finland; \{masoud.shokrnezhad, tarik.taleb\}@oulu.fi} \\
        \textsuperscript{2} \textit{Ruhr University Bochum, Bochum, Germany; tarik.taleb@rub.de} 
    }
}

\maketitle

\begin{abstract}
Anticipation for 6G's arrival comes with growing concerns about increased energy consumption in computing and networking. The expected surge in connected devices and resource-demanding applications presents unprecedented challenges for energy resources. While sustainable resource allocation strategies have been discussed in the past, these efforts have primarily focused on single-domain orchestration or ignored the unique requirements posed by 6G. To address this gap, we investigate the joint problem of service instance placement and assignment, path selection, and request prioritization, dubbed PIRA. The objective function is to maximize the system's overall profit as a function of the number of concurrently supported requests while simultaneously minimizing energy consumption over an extended period of time. In addition, end-to-end latency requirements and resource capacity constraints are considered for computing and networking resources, where queuing theory is utilized to estimate the Age of Information (AoI) for requests. After formulating the problem in a non-linear fashion, we prove its NP-hardness and propose a method, denoted ORIENT. This method is based on the Double Dueling Deep Q-Learning (D3QL) mechanism and leverages Graph Neural Networks (GNNs) for state encoding. Extensive numerical simulations demonstrate that ORIENT yields near-optimal solutions for varying system sizes and request counts.
\end{abstract}


\section{Introduction}\label{S_NTR}

The advent of the 6th generation of telecommunication systems (6G) signifies a pivotal era marked by unparalleled connectivity and technological advancements. With ultra-low End-to-End (E2E) latency (less than $1$ milisecond), exceeding $1$ terabit per second peak data rates, and ultra-high reliability surpassing $99.99999\%$ \cite{latva-aho_key_2020}, 6G promises to revolutionize industries such as holographic telepresence utilizing extended reality \cite{10293194}, dynamic metaverse empowered by semantic communications \cite{mazandarani2024semantic}, and quantum networking \cite{10299673}. However, achieving these capabilities raises substantial energy consumption concerns for both computing and networking resources. Presently, these resources consume around $200$ terawatt-hours of electricity annually, approximately $1\%$ of the global total \cite{yang_increasing_2022}. Many quality-sensitive applications may require uploading up to $50\%$ of data to computing facilities for processing \cite{noauthor_cisco_nodate}, adding even more strain to computing and networking resources. Moreover, the projected surge in 6G-connected devices and global data exacerbates the energy consumption challenge, underscoring the need for sustainable solutions.

In order to realize a 6G-enabled future, it may be necessary to create novel resource orchestration mechanisms to address impending energy challenges. The subject has been extensively studied in the literature. Xuan~\textit{et al.}~\cite{xuan_multi-agent_2023} addressed the Service Function Chaining (SFC) problem with the objective of minimizing energy consumption by proposing an algorithm based on multi-agent Reinforcement Learning (RL) and a self-adaptive division strategy. Solozabal~\textit{et al.}~\cite{solozabal_virtual_2020} investigated the same problem and proposed a single-agent solution. Other authors have also examined the SFC problem. By proposing a sampling-based Markov approximation method, Pham~\textit{et al.}~\cite{pham_traffic-aware_2020} solved the problem in an effort to minimize operational and traffic energy consumption. Santos~\textit{et al.}~\cite{santos_availability-aware_2021} developed two policy-aware RL algorithms based on actor-critic and proximal policy optimization to maximize availability while minimizing energy consumption. Reducing energy consumption was considered in the Service Function (SF) placement problem as well. Sasan \textit{et al.} \cite{sasan2024joint} presented a heuristic algorithm to tackle the joint problem of network slicing, path selection, and SF placement, with the objective of maximizing user acceptance while minimizing energy consumption. Farhoudi~\cite{farhoudi2023qos} and He~\textit{et al.} \cite{he_leveraging_2023} investigated a comparable problem and proposed RL-based solutions, taking into account the dynamic nature of service requests and overall cost considerations (including operation, deployment, and transmission), respectively.

While effective in specific contexts, the mentioned methods may not be suitable for 6G systems. These approaches prioritize energy efficiency over maximizing device support, whereas achieving an E2E efficient solution requires holistic management of computing and networking resources, considering the stringent Quality of Service (QoS) demands of 6G. Furthermore, certain studies overlook or oversimplify critical network parameters like latency, which contradicts the intricate and dynamic requirements of 6G systems. This paper addresses this gap by investigating the joint problem of allocating computing and networking resources (service instance placement and assignment, path selection, and request prioritization), termed PIRA. The objective is to optimize the system's overall profit (as a function of supported concurrent requests) while minimizing energy consumption over time, accounting for E2E latency and resource capacity constraints. The $M/M/1$ queuing model is employed to accurately evaluate request latency on compute nodes and network devices. To solve this problem, we propose ORIENT, an approach leveraging Double Dueling Deep Q-Learning (D3QL) reinforced by Graph Neural Networks (GNNs). This hybrid method effectively encodes the system state and facilitates the identification of near-optimal solutions.

The remainder of this paper is organized as follows. Section \ref{S_SST} introduces the system model. PIRA is defined and formulated in Section \ref{S_PRB}, and ORIENT is presented in Section \ref{S_ORN}. Finally, numerical results are illustrated in Section \ref{S_SML}, followed by concluding remarks in Section \ref{S_CNC}.

\section{System Model}\label{S_SST}
\begin{figure}[!t]
\centerline{\includegraphics[width=3.4in]{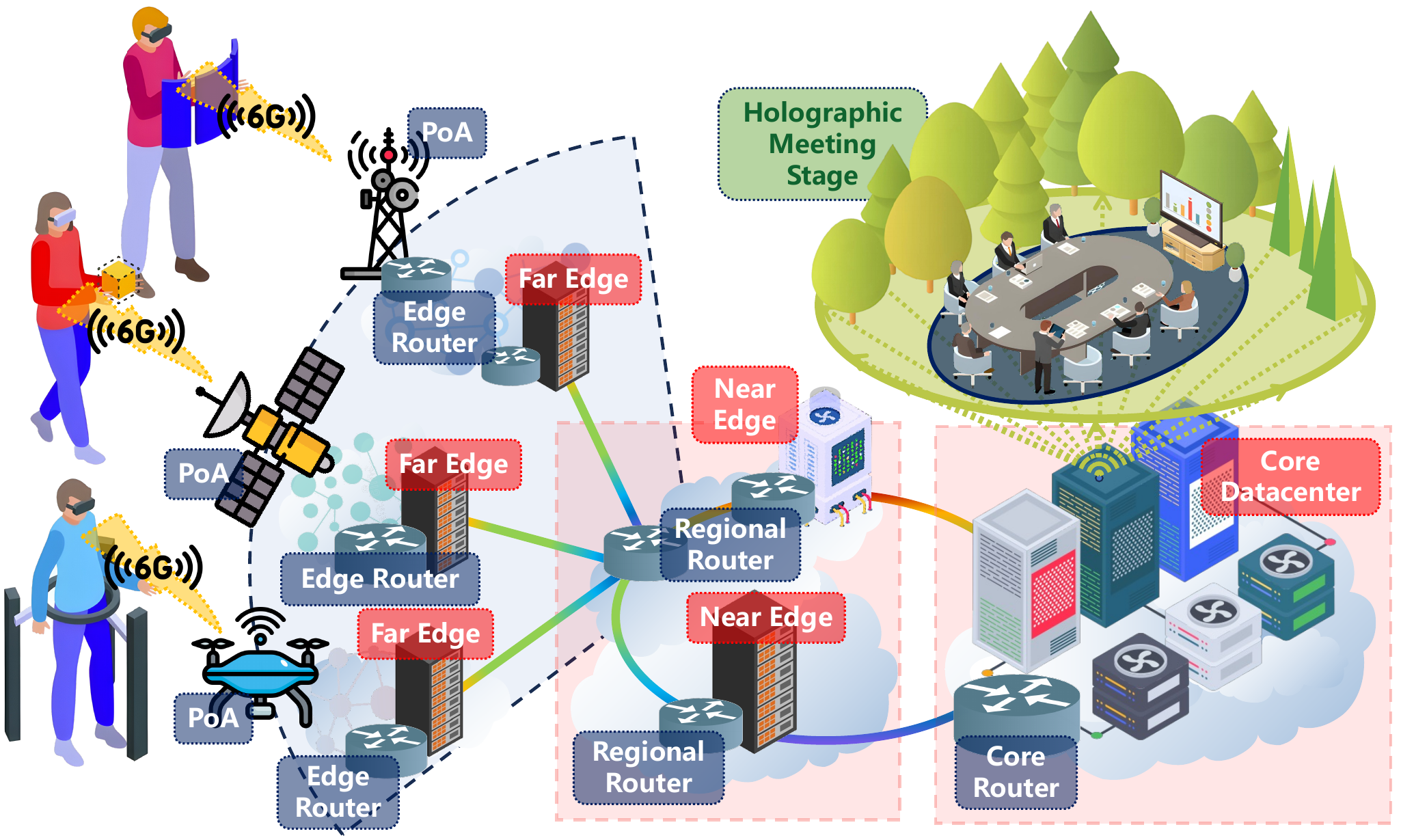}}
\caption{The system model, including network devices and distributed compute nodes facilitating holographic telepresence services for end users.}
\label{fig_system_model}
\vspace{-15pt}
\end{figure}

As shown in Fig. \ref{fig_system_model}, the following is an explanation of the two main components of the system: resources and requests.

\subsection{Resources}\label{SS_RSR}
The 6G system examined in this paper is an integrated infrastructure of computing and networking resources comprised of $\mathcal{N}$ network devices and $\mathcal{V}$ compute nodes (radio resources are excluded) \cite{10207694}. $\mathbb{N} = \{n | 0 \leq n \leq \mathcal{N}\}$ is the set of network devices, and $\mathbb{V} = \{v | 0 \leq v \leq \mathcal{V}\}$ denotes the set of compute nodes. Compute nodes are connected through network devices via $\mathcal{P}$ paths, the set of which is denoted by $\mathbb{P} = \{p | 0 \leq p \leq \mathcal{P}\}$, and the immediate network device of compute node $v$ is indicated by $n_v$. Each path $p$ contains a set of links $\mathbb{L}_p \subset \mathbb{L}$, where $\mathbb{L} = \{l:(n, n') | n, n' \in \mathbb{N}\}$ is the set of all network links, and $\mathcal{L}$ is its size. Network devices and compute nodes are priority-aware, i.e., $\mathbb{K} = \{k | 0 \leq k \leq \mathcal{K}\}$ is regarded as the set of permissible priority levels (where lower levels indicate higher priorities), and the resources in both domains are virtually partitioned, isolated, and guaranteed for each priority level $k$. Note that higher priorities receive a larger share of available resources than lower priorities.

To evaluate the performance of allocated resources, we will employ the $M/M/1$ queuing model for each priority level on network devices and compute nodes assuming that this theory's stability requirements are met and that all queues are independent. The service rate allocated to priority level $k$ on network device $n$ is $\widehat{\mathcal{B}_{n,k}}$, and those packets leaving this queue will be forwarded through their corresponding link, let's call it $l$, allocating $\widehat{\mathcal{B}_{l,k}}$ bandwidth. Note that the overall capacity of this network device and link is constrained by $\widehat{\mathcal{B}_n}$ and $\widehat{\mathcal{B}_l\wdot}$, respectively. Similarly, the requests of priority level $k$ will be served on compute node $v$ leveraging a queue with dedicated service rate $\widehat{\mathcal{C}_{v,k}}$, and this node is equipped with computing resources limited to a predefined capacity threshold, dubbed $\widehat{\mathcal{C}_v}$. In addition, the energy consumption for transmitting bandwidth units over network device $n$ is $\widetilde{\mathcal{E}_n}$, and compute node $v$ consumes $\widetilde{\mathcal{E}_v}$ energy per capacity unit and $\overline{\mathcal{E}_v}$ energy when its state changes (from the idle mode to the operation mode or vice versa).

\subsection{Requests}\label{SS_RQS}
This paper investigates the system for $\mathcal{T}$ time slots while a set of $\mathcal{R}_t$ requests, denoted by $\mathbb{R}_t=\{r|0 \leq r \leq \mathcal{R}_t\}$, arrives at time slot $t \in \mathbb{T}=\{t|1 \leq t \leq \mathcal{T}\}$. The set of all requests is $\mathbb{R}=\{\mathbb{R}_t|1 \leq t \leq \mathcal{T}\}$, and $\mathcal{R}$  represents the number of all requests. Each request $r$ enters the system through an edge network device, denoted by $n_r$ and referred to as its Point of Arrival (PoA), and orders service $s_r$ from the set of obtainable service instances, that is $\mathbb{I}=\{i|0 \leq i \leq \mathcal{I}\}$, where instance $i$ provides service $s_i$. In order to successfully fulfill each request, one instance of its target service must be replicated on one of the compute nodes in order to receive the request, process it, and return it to its entry point so that it can be delivered to the end user. $\widehat{\mathcal{C}_{i\wdot}}$ represents the maximum capacity of instance $i$. To fulfill each user's request, its QoS requirements must be met, including the minimum service capacity and network bandwidth, as well as the maximum tolerable E2E latency,  denoted by $\widecheck{\mathcal{C}_{r\wdot}}$, $\widecheck{\mathcal{B}_r}$, and $\widecheck{\mathcal{D}_r}$, respectively. Besides, the maximum permissible packet size for request $r$ is $\widehat{\mathcal{H}_r}$. If request $r$ is successfully completed, the system will achieve a profit, that is $\gamma_r$. Note that the arrival rate for each queue is determined by a Poisson process and is assumed to be the sum of $\widecheck{\mathcal{C}_{r\wdot}}$ (for compute queues) and $\widecheck{\mathcal{B}_r}$ (for network queues) for all requests assigned to that queue, respectively.

\section{Problem Definition}\label{S_PRB}
This section discusses the joint problem of instance placement and assignment, request prioritization, and path selection for integrated compute-network infrastructures to maximize the overall profit of the system while minimizing its energy consumption. In this section, the constraints and objective function are formulated, followed by the problem statement as a Mixed-Integer Non-Linear Programming (MINLP) formulation and its complexity analysis.

\subsection{Instance Orchestration Constraints}\label{SS_NST}
Constraints C1-C6 assign requests to instances and place them on compute nodes while maintaining the capacity constraints of instances and compute nodes. Considering that $\ddot{\mathcal{I}}^{t}_{r,i}$ is a binary variable whose value is $1$ if request $r$ is assigned to instance $i$ at time slot $t$, C1 ensures that each request $r$ is assigned to no more than one instance of its service for each time slot $t$. C2 defines a new binary variable, $\dot{\mathcal{I}}^{t}_{i}$, which indicates that whether instance $i$ is activated at time slot $t$. If $\sum_{\mathbb{R}_{t}} \ddot{\mathcal{I}}^{t}_{r,i}$ is equal to or greater than $1$ (i.e., at least one request is assigned to instance $i$), $(\sum_{\mathbb{R}_{t}} \ddot{\mathcal{I}}^{t}_{r,i})/\mathcal{R}_{t}$ will be a small number (between $0$ and $1$) and $\sum_{\mathbb{R}_{t}} \ddot{\mathcal{I}}^{t}_{r,i}$ will be a large number, so $\dot{\mathcal{I}}^{t}_{i}$ will be set to $1$. Otherwise, both sides of the equation will equal $0$, causing $\dot{\mathcal{I}}^{t}_{i}$ to also equal $0$. C3 ensures that each activated instance is assigned to exactly one compute node, where $\ddot{\mathcal{G}}^{t}_{i,v}$ is a binary variable indicating the compute node of instance $i$ at time slot $t$. Similar to C2, C4 defines $\dot{\mathcal{G}}^{t}_{v}$ as a binary variable to determine whether compute node $v$ should be activated at time slot $t$. Then, it must be assured that assigned requests do not exceed the capacity limitations of instances and compute nodes (C5 and C6).

\vspace{-10pt}
\footnotesize 
\begin{align}\label{CS_NST}
    &
    \sum\nolimits_{\mathbb{I} | s_i = s_r } \ddot{\mathcal{I}}^{t}_{r,i} 
    \leq 
    1 
    \quad 
    \forall t, r 
    \in 
    \mathbb{T}, \mathbb{R}_{t}
    \textcolor{white}{ttttttttttttttttttttttttttttttttttt}
    \tag{C1} \\
    & 
    \frac{1}{\mathcal{R}_{t}} \cdot \sum\nolimits_{\mathbb{R}_{t}} \ddot{\mathcal{I}}^{t}_{r,i} 
    \leq 
    \dot{\mathcal{I}}^{t}_{i} 
    \leq 
    \sum\nolimits_{\mathbb{R}_{t}} \ddot{\mathcal{I}}^{t}_{r,i}
    \quad 
    \forall t, i 
    \in 
    \mathbb{T}, \mathbb{I}
    \tag{C2} \\
    &
    \sum\nolimits_{\mathbb{V}} \ddot{\mathcal{G}}^{t}_{i,v} 
    = 
    \dot{\mathcal{I}}^{t}_{i}
    \quad 
    \forall t, i 
    \in 
    \mathbb{T}, \mathbb{I}  
    \tag{C3} \\
    &
    \frac{1}{\mathcal{I}} \cdot \sum\nolimits_{\mathbb{I}} \ddot{\mathcal{G}}^{t}_{i,v} 
    \leq 
    \dot{\mathcal{G}}^{t}_{v} 
    \leq 
    \sum\nolimits_{\mathbb{I}} \ddot{\mathcal{G}}^{t}_{i,v}
    \quad 
    \forall t, v 
    \in 
    \mathbb{T}, \mathbb{V}
    \tag{C4} \\
    &
    \sum\nolimits_{\mathbb{R}_t} \widecheck{\mathcal{C}_r} \cdot \ddot{\mathcal{I}}^{t}_{r,i} 
    \leq 
    \widehat{\mathcal{C}_{i\wdot}}
    \quad
    \forall t, i 
    \in 
    \mathbb{T}, \mathbb{I} 
    \tag{C5} \\
    &
    \sum\nolimits_{\mathbb{I}} \widehat{\mathcal{C}_{i\wdot}} \cdot \ddot{\mathcal{G}}^{t}_{i,v} 
    \leq 
    \widehat{\mathcal{C}_v}
    \quad
    \forall v, t 
    \in 
    \mathbb{V}, \mathbb{T} 
    \tag{C6}
\end{align}
\normalsize

\subsection{Path Selection Constraints}\label{SS_PTH}
Constraints C7-C9 ensure that an E2E path is selected for each request considering the capacity constraints of network links and the traffic pattern, where packets of each request enter the network through its PoA and, after visiting its assigned instance, are returned to the same PoA to be handed off to the corresponding end user. C7 determines the allocated path for each request $r$, ensuring that it originates and terminates at $n_r$ and traverses the network device directly connected to the compute node hosting the instance assigned to the request. In this constraint, $\ddot{f}^{t}_{r,p}$ is a binary variable that represents the assigned path of request $r$ at time slot $t$. Finally, C8 and C9 maintain the maximum capacity of network links and devices.

\vspace{-10pt}
\footnotesize 
\begin{align}\label{CS_PTH}
    &
    \sum\nolimits_{\mathbb{P} |  n_r \& n_v \in p } \ddot{f}^{t}_{r, p}
    = 
    \ddot{\mathcal{I}}^{t}_{r,i} \cdot \ddot{\mathcal{G}}^{t}_{i,v}
    \quad
    \forall t, r, i, v, 
    \in 
    \mathbb{T}, \mathbb{R}, \mathbb{I}, \mathbb{V}
    \tag{C7} \\
    &
    \sum\nolimits_{\mathbb{R}_{t}, \mathbb{P} | l \in \mathbb{L}_p } \widecheck{\mathcal{B}_r} \cdot \ddot{f}^{t}_{r,p} 
    \leq 
    \widehat{\mathcal{B}_l\wdot}
    \quad
    \forall t, l 
    \in 
    \mathbb{T}, \mathbb{L}
    \textcolor{white}{tttttttttttttttttttttttttt}
    \tag{C8} \\
    &
    \sum\nolimits_{\mathbb{R}_{t}, \mathbb{P} | n \in \mathbb{L}_p } \widecheck{\mathcal{B}_r} \cdot \ddot{f}^{t}_{r,p} 
    \leq 
    \widehat{\mathcal{B}_n\wdot}
    \quad
    \forall t, n 
    \in 
    \mathbb{T}, \mathbb{N}
    \tag{C9} 
\end{align}
\normalsize

\subsection{Request Prioritization Constraints}\label{SS_PRR}
\vspace{-10pt}
\footnotesize 
\begin{align}\label{CS_PRR1}
    &
    \sum\nolimits_{\mathbb{K}} \ddot{\varrho}^{t}_{r,k} 
    = 
    \sum\nolimits_{\mathbb{I}} \ddot{\mathcal{I}}^{t}_{r,i}
    \quad 
    \forall t, r
    \in 
    \mathbb{T}, \mathbb{R}_{t}
    \textcolor{white}{ttttttttttttttttttttttttttttttt}
    \tag{C10} \\
    &
    \sum\nolimits_{\mathbb{R}_{t}, \mathbb{I}} \widecheck{\mathcal{C}_r} \cdot \ddot{\varrho}^{t}_{r,k} \cdot  \ddot{\mathcal{I}}^{t}_{r,i} \cdot \ddot{\mathcal{G}}^{t}_{i, v}
    <
    \widehat{\mathcal{C}_{v,k}} 
    \quad
    \forall t, k, v 
    \in 
    \mathbb{T}, \mathbb{K}, \mathbb{V}
    \tag{C11} \\
    &
    \sum\nolimits_{\mathbb{R}_{t}, \mathbb{P} | l \in \mathbb{L}_p } \widecheck{\mathcal{B}_r} \cdot \ddot{\varrho}^{t}_{r,k} \cdot \ddot{f}^{t}_{r,p}  
    < 
    \widehat{\mathcal{B}_{l,k}\wdot}
    \quad
    \forall t, k, l 
    \in
    \mathbb{T}, \mathbb{K}, \mathbb{L}
    \tag{C12} \\
    &
    \sum\nolimits_{\mathbb{R}_{t}, \mathbb{P} | n \in \mathbb{L}_p } \widecheck{\mathcal{B}_r} \cdot \ddot{\varrho}^{t}_{r,k} \cdot \ddot{f}^{t}_{r,p}  
    <
    \widehat{\mathcal{B}_{n,k}\wdot}
    \quad
    \forall t, k, n 
    \in 
    \mathbb{T}, 
    \mathbb{K}, \mathbb{N}
    \tag{C13}
\end{align}
\normalsize

To maintain integrity, it's crucial to prevent any overuse of resources allocated to each priority level. Given that $\ddot{\varrho}^{t}_{r,k}$ is the priority of request $r$ at time slot $t$, C10 promises that the request's priority is determined if an instance is assigned to serve it. Then, C11 to C13 satisfy the capacity constraints of priority queues on compute nodes and network resources.

\subsection{Latency Constraints}\label{SS_LTN}
Each packet has to wait for three sources of latency through its request's assigned E2E path in the system: 1) the service latency experienced at the network devices included in the path, 2) the transmission latency over the network links of the path, and 3) the service latency at the assigned compute node. Since the average latency of a packet in a $M/M/1$ queue is equal to $1/(\mu - \lambda )$ when the arrival rate is $\lambda$ and the service rate is $\mu$, the average latency experienced by the packets of request $r$ at network device $n$ allocated to priority level $k$ during time slot $t$ can be calculated as C14. In this constraint, the numerator will be $0$ for network devices and priority levels that have not been allocated to request $r$, causing $\ddot{\mathcal{D}}^{t}_{r, n,k}$ to equal $0$. Otherwise, the numerator will be $1$, and the latency will be determined following the adopted queuing theorem with the arrival rate of the queue set to the overall bandwidth of requests assigned to priority level $k$ and traversing network device $n$. C15 follows the same logic to calculate the average latency of request $r$ allocated to priority level $k$ at time slot $t$ on compute node $v$. C16 calculates the transmission latency of request $r$ over link $l$ at time slot $t$, considering its priority level and maximum packet size, if the link is part of the path assigned to the request. Otherwise, the latency will be $0$. Finally, C17 determines the Age of Information (AoI), followed by C18, which ensures the maximum acceptable latency requirement of requests. 

\vspace{-10pt}
\footnotesize 
\begin{align}\label{CS_LTC}
    &
    \ddot{\mathcal{D}}^{t}_{r,n,k} 
    = \frac
    {
        \sum\nolimits_{\mathbb{P} | n \in \mathbb{L}_p }  \ddot{\varrho}^{t}_{r,k} \cdot \ddot{f}^{t}_{r,p}
    }
    {
        \widehat{\mathcal{B}_{n,k}} - \sum\nolimits_{\mathbb{R}_{t}, \mathbb{P} | n \in \mathbb{L}_p } \widecheck{\mathcal{B}_{r'}} \cdot \ddot{\varrho}^{t}_{r',k} \cdot \ddot{f}^{t}_{r',p}
    }
    \quad
    \begin{aligned}
        & \forall t, r, k, n \in \\
        & \mathbb{T}, \mathbb{R}_{t}, \mathbb{K},
        \mathbb{N} \\
    \end{aligned} 
    \tag{C14} \\
    &
    \ddot{\mathcal{D}}^{t}_{r,v,k} 
    = \frac
    {
        \sum\nolimits_{\mathbb{I}} \ddot{\varrho}^{t}_{r,k} \cdot \ddot{\mathcal{I}}^{t}_{r,i} \cdot \ddot{\mathcal{G}}^{t}_{i,v}
    }
    {
        \widehat{C_{v,k}} - \sum\nolimits_{\mathbb{R}_{t}, \mathbb{I}} \widecheck{\mathcal{C}_{r'}} \cdot \ddot{\mathcal{I}}^{t}_{r',i} \cdot \ddot{\mathcal{G}}^{t}_{i,v}
    }
    \quad
    \begin{aligned}
        & \forall t, r, k, v \in \\
        & \mathbb{T}, \mathbb{R}_{t}, \mathbb{K},
        \mathbb{V} \\
    \end{aligned}
    \tag{C15} \\
    &
    \ddot{\mathcal{D}}^{t}_{r,l,k} 
    = 
    \frac
    {\widehat{\mathcal{H}_r}}
    {\widehat{\mathcal{B}_{l,k}}} \cdot \sum\nolimits_{\mathbb{P} | l \in \mathbb{L}_p }  \ddot{\varrho}^{t}_{r,k} \cdot \ddot{f}^{t}_{r,p}
    \quad
    \forall t, r, l, k 
    \in 
    \mathbb{T}, \mathbb{R}_{t}, \mathbb{L},
    \mathbb{K}
    \textcolor{white}{tttt}
    \tag{C16} \\
    &
    \ddot{\mathcal{D}}^{t}_{r}
    = 
    \sum\nolimits_{\mathbb{N}, \mathbb{K}, \mathbb{V}, \mathbb{L}} ( \ddot{\mathcal{D}}^{ t}_{r,n,k} + \ddot{\mathcal{D}}^{t}_{r,v,k} + \ddot{\mathcal{D}}^{t}_{r,l,k} )
    \quad
    \forall t, r 
    \in 
    \mathbb{T}, \mathbb{R}_{t} 
    \tag{C17} \\
    &
    \ddot{\mathcal{D}}^{t}_{r}
    \leq
    \widecheck{\mathcal{D}_r}
    \quad
    \forall t, r
    \in 
    \mathbb{T}, \mathbb{R}_{t}
    \tag{C18}
\end{align}
\normalsize

\subsection{Objective Function}\label{SS_BJC}
The objective is to maximize the overall profit while minimizing the energy consumption of resources, that is:

\vspace{-10pt}
\footnotesize 
\begin{align}\label{CS_BJC}
    &
    \sum\nolimits_{\mathbb{T}, \mathbb{R}_t} ( \gamma_r \cdot \sum\nolimits_{\mathbb{I}} \ddot{\mathcal{I}}^{t}_{r,i} ) - \alpha \cdot ( \sum\nolimits_{\mathbb{N}} \ddot{\mathcal{E}}_n + \sum\nolimits_{\mathbb{V}} \ddot{\mathcal{E}}_v ), \textcolor{white}{ttttttttttttt}
    \tag{OF}
    \textcolor{white}{ttt}
\end{align}
\normalsize
where $\sum_{\mathbb{I}} \ddot{\mathcal{I}}^{t}_{r,i}$ is $1$ if request $r$ is supported at time slot $t$, $\alpha$ is a small positive number, and $\ddot{\mathcal{E}}_v$ and $\ddot{\mathcal{E}}_n$ represent, respectively, the total energy consumption of compute node $v$ and network device $n$. Note that $\alpha$ must be set such that the total profit exceeds the total amount of energy consumed. Otherwise, supporting requests would result in a negative objective function value, and the only optimal solution would be to support no requests, making \ref{CS_BJC} equal to $0$. To determine $\ddot{\mathcal{E}}_n$, where the only source of energy consumption is transmitting requests' data, the following equations are employed:

\vspace{-7pt}
\footnotesize 
\begin{align}\label{CS_NTN}
    &
    \ddot{\mathcal{E}}_n 
    = \widetilde{\mathcal{E}_n} \cdot \sum\nolimits_{\mathbb{T}, \mathbb{R}_{t}, \mathbb{P} | n \in \mathbb{L}_p } \widecheck{\mathcal{B}_{r}} \cdot \ddot{f}^{t}_{r,p}
    \textcolor{white}{ttttttttttttttttttttttttttttttttttt}
\end{align}
\normalsize

To calculate {\small$\ddot{\mathcal{E}}_v$}, it should be noted that the energy consumed on each compute node has two primary sources: 1) the energy consumed to service each unit of requests' data, and 2) the energy consumed during booting up or shutting down the compute node. Consequently, $\ddot{\mathcal{E}}_v$ for each $v \in \mathbb{V}$ is:

\vspace{-7pt}
\footnotesize 
\begin{align}\label{CS_CNN}
    \widetilde{\mathcal{E}_v} \cdot \sum\nolimits_{\mathbb{T}, \mathbb{R}_{t}, \mathbb{I}} \widecheck{\mathcal{C}_{r}} \cdot \ddot{\mathcal{I}}^{t}_{r,i} \cdot \ddot{\mathcal{G}}^{t}_{i,v} 
    + 
    \overline{\mathcal{E}_v} \cdot \sum\nolimits_{\mathbb{T} | 0 \leq t<\mathcal{T}} \dot{\mathcal{G}}^{t}_{v} \oplus \dot{\mathcal{G}}^{t+1}_{v}
\textcolor{white}{ttttttttt}
\end{align}
\normalsize
In this equation, $\ddot{\mathcal{I}}^{t}_{r,i} \cdot \ddot{\mathcal{G}}^{t}_{i,v}$ equals $1$ if compute node $v$ is selected as the host for request $r$; therefore, the first part of the equation calculates the total energy consumption of compute node $v$ to service its assigned requests. Assuming $\dot{\mathcal{G}}^{0}_{v}$ equals $0$, $\dot{\mathcal{G}}^{t}_{v} \oplus \dot{\mathcal{G}}^{t+1}_{v}$ is equal to $1$ in the second part if and only if $\dot{\mathcal{G}}^{t}_{v}$ and $\dot{\mathcal{G}}^{t+1}_{v}$ differ, representing a boot-up or shutdown for compute node $v$. Then, the second part demonstrates the total energy consumption of state transitions in compute note $v$ within $\mathbb{T}$.

\subsection{Problem}\label{SS_PRB}
Considering the constraints and the objective function, the Problem of Integrated Resource Allocation (PIRA) is:
\begin{equation}\label{EQ_PIRA}
    \footnotesize \text{PIRA: } \max \text{ OF} \text{ s.t. } \text{C1 - C18.} 
    \textcolor{white}{tttttttttttttttttttttttttttttttttttttttttt}
\end{equation}
The optimal solution involves assigning requests with stringent latency requirements to high-priority queues  and keeping those queues as empty as possible to minimize latency. Resource selection should aim to minimize activated resources and consider future requests to reduce energy consumption through fewer start-ups and shutdowns, improving energy efficiency.

\subsection{Complexity Analysis}\label{SS_CMP}
The problem defined in (\ref{EQ_PIRA}) is an extended version of the Multi-Dimensional Knapsack (MDK) problem. Assume the problem is relaxed and reformulated specifically for time slot $t$ as the problem of maximizing profit and minimizing energy consumption while the only decision is to assign requests to instances concerning only their capacity constraints. Since the MDK problem is NP-hard and this relaxed version is an MDK problem with at least $\mathcal{R}_t$ items and $\mathcal{I}$ knapsacks, PIRA is at least as difficult as the MDK problem and is also NP-hard. 

\section{ORIENT}\label{S_ORN}

This section proposes an RL-based priORIty-aware Energy-efficieNt laTency-sensitive resource allocation approach (ORIENT) to find near-optimal solutions for PIRA. Subsequently, the learning mechanism is elaborated upon, followed by an explanation of the agent's design, and concluding with a description of the algorithm.

\subsection{Learning Mechanism}
Given the continuous operation of the system defined in this paper and the recurring necessity for consistent resource allocation decisions at each time slot of PIRA, the adoption of RL presents itself as a viable means to enhance decision-making proficiency to solve it. Within the framework of RL, an agent undergoes a process of learning by means of trial and error at each step (here, time slot), with the primary aim of optimizing a specific decision-making problem. The system's designer defines a reward function in alignment with the objectives of the problem. By learning and following the optimal strategy derived from this reward function, the agent aims to maximize cumulative discounted rewards, regardless of the initial state. Among various RL-based algorithms, Q-Learning stands out as widely acknowledged.

In Q-Learning, every state-action pair is associated with a numeric value referred to as the Q-value, where the agent selects the action with the maximum Q-value at each step. In Deep Q-Learning (DQL), a Deep Neural Network (DNN) serves as the approximator for these Q-values. In this arrangement, the state and action are presented as inputs, and the DNN-based $Q$-function encompassing all feasible actions, denoted by $Q(s, .; \boldsymbol{\mathcal{W}})$, is generated as the output and systematically updated over time according to the following equation:
\begin{equation}\label{eq_DQL_bellman}
    \footnotesize \boldsymbol{\mathcal{W}}^{t+1} = \boldsymbol{\mathcal{W}}^t + \sigma[Y^t - Q(\boldsymbol{S}^t, a^t; \boldsymbol{\mathcal{W}}^t)]\nabla_{\boldsymbol{\mathcal{W}}^t} \cdot Q(\boldsymbol{S}^t, a^t; \boldsymbol{\mathcal{W}}^t)
    \textcolor{white}{tttttttt}
\end{equation}
In this equation, $\boldsymbol{\mathcal{W}}$ is the set of DNN weights, $\sigma$ is a scalar step size, $\boldsymbol{S}^t$ and $a^t$ are the agent's state and action at time slot $t$, and $Y^t$ (also known as the target) shows the maximum value expected to be achieved by following $a^t$ at $\boldsymbol{S}^t$. $Y^t$ is the only variable that must be estimated in this equation, and in Double DQL (DDQL), where the selection and evaluation processes are decoupled, it can be expressed as follows:
\begin{equation}\label{eq_DDQL_target}
    \footnotesize Y^t  = r^{t+1} + \gamma \; \widehat{Q}(\boldsymbol{S}^{t+1}, a', \boldsymbol{\mathcal{W}}^{t-}),
    \textcolor{white}{tttttttttttttttttttttttttttttttttttt}
\end{equation}
where $r^{t+1}$ is the earned reward at time slot $t+1$, $\gamma \in [0,1]$ is a discount factor that balances the importance of immediate and future rewards, $a' = \text{argmax}_{a \in \boldsymbol{\mathcal{A}}} Q(\boldsymbol{S}^{t+1}, a, \boldsymbol{\mathcal{W}}^t)$, and $\boldsymbol{\mathcal{A}}$ is the set of actions. In this equation, $\boldsymbol{\mathcal{W}}$ represents the set of weights for the main $Q$ and is updated in each step, whereas $\boldsymbol{\mathcal{W}}^-$ is for the target $\widehat{Q}$ and is replaced with the weights of the main network every $t$ steps. In other words, $\widehat{Q}$ remains a periodic copy of $Q$.

Furthermore, we augment DDQL by incorporating the dueling concept introduced by Wang \textit{et al.} \cite{wang2016dueling}. Unlike conventional DDQL, which directly approximates Q-values using DNNs, this method initially computes separate estimators for state values ($\psi$) and action advantages ($\varphi$). Q-values are then derived from these estimators, as illustrated below:
\begin{equation}
\label{eq_dueling}
\footnotesize Q(\boldsymbol{S}^t, a^t, \boldsymbol{\mathcal{W}}^t) = \psi(\boldsymbol{S}^t, \boldsymbol{\mathcal{W}}^t)  \Bigg( \varphi(\boldsymbol{S}^t, a^t, \boldsymbol{\mathcal{W}}^t) - \frac{\Phi}{\left| \boldsymbol{\mathcal{A}} \right|} \Bigg)
\textcolor{white}{ttttttttttttt}
\end{equation}
where $\Phi = \sum_{\boldsymbol{\mathcal{A}}}^{}\varphi(\boldsymbol{S}^t, a^{\prime}, \boldsymbol{\mathcal{W}}^t)$. The primary benefit is the ability to generalize learning across actions without modifying the learning algorithm, which improves policy evaluation in the presence of numerous actions with similar state values. As a result of combining the Dueling technique and DDQL, we can expect that the resultant D3QL agent will outperform its predecessors. To bolster the effectiveness and resilience of D3QL, observed transitions are archived in a memory bank known as the experience memory. The learning process entails randomly selecting transitions from this repository \cite{mnih_human-level_2015}.

\subsection{Agent Customization}
The first step toward exploiting D3QL to solve PIRA is to define the agent's action space, state space, and reward.
\subsubsection*{Action Space} 
We define the action space as set $\boldsymbol{\mathcal{A}} = \{a: (i, v, p, k) | i, v, p, k \in \mathbb{I}, \mathbb{V}, \mathbb{P}, \mathbb{K}\}$. During each time slot and for every request, a specific action must be executed to finalize the resource allocation pertaining to that request.
\subsubsection*{State Space} 
For encoding the system's state, an architecture involving aggregation GNN layers is employed, constructing an aggregation sequence across all compute nodes, iteratively facilitating information exchange with neighboring nodes. Therefore, at time slot $t$ when request $r$ is on the verge of receiving service, the system state is denoted as $\boldsymbol{S}^t(r) = \{\boldsymbol{S}^t_\mathbb{V}(r), \boldsymbol{S}^t_\mathbb{P}(r)\}$ and can be formally defined as:

\vspace{-10pt}
\footnotesize 
\begin{align}\label{state_space}
    &
    \boldsymbol{S}^t_\mathbb{V}(r)
    = 
    \Bigg\{\Big[ [\widehat{\mathcal{C}^t_{v,k}} - \widecheck{\mathcal{C}_{r\wdot}}]_{\mathbb{K}}, [\ddot{\mathcal{D}}^{t}_{r,v,k} - \widecheck{\mathcal{D}_{r\wdot}}]_{\mathbb{K}},  \widetilde{\mathcal{E}_v}, \overline{\mathcal{E}_v} \cdot (1-\dot{\mathcal{G}}^{t}_{v})   \Big]_{\mathbb{V}} \Bigg\},
    \\
    &
    \boldsymbol{S}^t_\mathbb{P}(r)
    = 
    \Bigg\{\Big[ \big[\wedge_{\mathbb{L}_p} - \widecheck{\mathcal{B}_{r\wdot}} \big]_{\mathbb{K}}, [\ddot{\mathcal{D}}^{t}_{r} -\ddot{\mathcal{D}}^{t}_{r,v,k} - \widecheck{\mathcal{D}_{r\wdot}}]_{\mathbb{K}},  \widetilde{\mathcal{E}_n} \Big]_{\mathbb{P}} \Bigg\},
    \textcolor{white}{ttttttttt}
\end{align}
\normalsize
where $\wedge_{\mathbb{L}_p} = \min\nolimits_{\mathbb{L} | n, l \in \mathbb{L}_p} \{\widehat{\mathcal{B}^t_{n,k}}, \widehat{\mathcal{B}^t_{l,k}}\}$. $\boldsymbol{S}^t_\mathbb{V}(r)$ and $\boldsymbol{S}^t_\mathbb{P}(r)$ represent the embeddings of compute nodes and network paths, respectively, which function as inputs for the GNN layers. These embeddings encompass the remaining resource capacity and the anticipated latency when request $r$ is allocated to them, as well as their associated energy consumption.

\subsubsection*{Reward}
Since the agent is designated to maximize OF, the reward should be engineered to reinforce the support of high-profit requests while selecting resources with low energy consumption. This goal is satisfied in \eqref{reward}, that is:
\begin{equation} \label{reward}
\footnotesize 
r^{t + 1} = 
\left\{\begin{array}{ll}
    0 & \mbox{otherwise} \\
    \dfrac{\mathcal{M}}{\text{OF}\big(\boldsymbol{S}^t(r),a^t\big) - \min_{\boldsymbol{\mathcal{A}}} \text{OF}\big(\boldsymbol{S}^t(r),a'\big)} &  r \text{ is met}
    \textcolor{white}{t}
\end{array}\right. 
\normalsize
\end{equation}
where $\mathcal{M} = \max_{\boldsymbol{\mathcal{A}}} \text{OF}\big(\boldsymbol{S}^t(r),a'\big) - \min_{\boldsymbol{\mathcal{A}}} \text{OF}\big(\boldsymbol{S}^t(r),a'\big)$, $\max\text{/}\min_{\boldsymbol{\mathcal{A}}} \text{OF}\big(\boldsymbol{S}^t(r),a'\big)$ is the maximum/minimum profit that can be achieved by allocating the available resources at time slot $t$ to request $r$ without considering any constraints or requirements, and $\text{OF}\big(\boldsymbol{S}^t(r),a^t\big)$ is the profit of the allocation provided by the agent. 
If the action fails to meet the requirements of the request, it results in a reward of $0$. Conversely, actions that yield greater profits correspond to higher rewards.

\begin{table}[t!]
\caption{Simulation Parameters.}
\begin{center}
\begin{tabular}{|c|c|}
\hline
\textbf{Parameter} & \textbf{Value} \\
\hline
number of priority levels & $4$ \\
resource capacity bounds & $\sim \mathcal{U}\{250, 300\}$ mbps \\
Instance capacity bound & $20$ mbps \\
energy consumptions per capacity unit & $\sim \mathcal{U}\{10, 20\}$ \\
energy consumptions per state transition & $\sim \mathcal{U}\{100, 200\}$  \\
capacity requirement per request & $\sim \mathcal{U}\{4, 8\}$ mbps\\
bandwidth requirement per request & $\sim \mathcal{U}\{2, 10\}$ mbps\\
latency requirement per request & $\sim \mathcal{U}\{1, 3\}$ ms\\
packet size per request & $1$ \\
profit per request ($\gamma_r$) & $\mathcal{U}\{5, 15\}$ \\
\hline
\end{tabular}
\label{T_SML}
\end{center}
\vspace{-15pt}
\end{table}

\begin{algorithm}[]\label{ORIENT}
\KwInput{$\mathcal{T}$, $\epsilon'$, and $\widetilde{\epsilon}$}
$\boldsymbol{\Omega} \leftarrow \emptyset$, $\boldsymbol{\mathcal{W}} \leftarrow \mathbf{0}$, $\boldsymbol{\mathcal{W}^-} \leftarrow \mathbf{0}$, $\epsilon \gets 1$, $memory \gets \{\}$\\
\For{each $t$ in $[0:\mathcal{T}]$}
{
    \If{new request $r$ is arrived}
    {
        calculate $\boldsymbol{S}^t(r) = \{\boldsymbol{S}^t_\mathbb{V}(r), \boldsymbol{S}^t_\mathbb{P}(r)\}$ \\
        $\zeta \gets$ generate a random number from $[0:1]$ \\
        \If{$\zeta > \epsilon$}
        {
            $a^t = (i, v, p , k) \gets$ argmax$_{\boldsymbol{\mathcal{A}}} Q(\boldsymbol{S}^t, a^{\prime}, \boldsymbol{\mathcal{W}}^t)$ \\
        }
        \Else
        {
            select a random $a^t = (i, v, p , k)$ from $\boldsymbol{\mathcal{A}}$
        }
        calculate $r^{t+1}$ \\
        \If{$r^{t+1} > 0$}
        {
            Establish request $r$ connection based on $a^t$ \\
        }
        $memory \gets memory \cup \{(\boldsymbol{S}^t, a^t, r^{t+1})\}$ \\
        choose a batch of samples from $memory$\\
        train the agent\\
        \If{$\epsilon > \widetilde{\epsilon}$}
        {
            $\epsilon \gets \epsilon - \epsilon'$
        }
        $\boldsymbol{\Omega} \leftarrow \boldsymbol{\Omega} \cup \{(t, r, a^t)\}$
    }
}
return $\boldsymbol{\Omega}$
\caption{ORIENT}
\end{algorithm}

\subsection{ORIENT's Algorithm}
ORIENT is detailed in Algorithm \ref{ORIENT}. In this algorithm, $\epsilon'$ and $\widetilde{\epsilon}$ are small positive integers that control the $\epsilon$-greedy mechanism. During each time slot $t$, the agent receives notifications of new request arrivals ($r$), and it computes the state based on request $r$ requirements and the current system state. The action is then chosen using an $\epsilon$-greedy policy, which follows the evaluation function of the corresponding agent with probability $(1-\epsilon)$ and selects a random action with probability $\epsilon$. Subsequently, the reward is calculated, and if it exceeds $0$, it indicates that $a^t$ is feasible and meets request $r$'s QoS requirements, enabling its connection based on $a^t$ allocations. Finally, the experience memory is updated, samples are drawn from the memory bank, and the agent undergoes training. during the training process, $\epsilon$ decreases from $1$ to $\widetilde{\epsilon}$. The algorithm yields $\boldsymbol{\Omega}$ as the history of allocations.

\section{Performance Evaluation}\label{S_SML}

\begin{figure*}[t!]\centering
\includegraphics[width=7.15in]{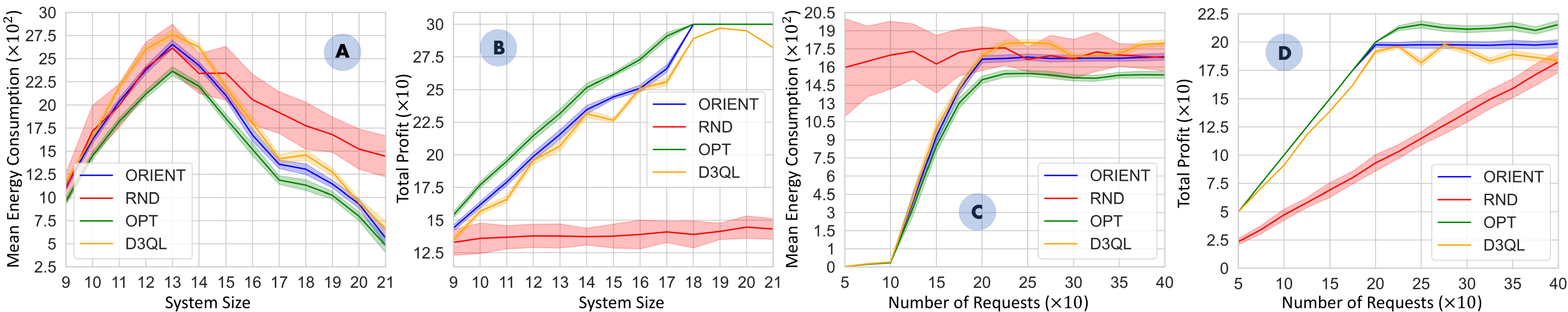}
\vspace{-15pt}
  \caption{The mean energy consumption of supported requests and the total profit vs. the system size (A \& B) and the number of all requests (C \& D).}
  \label{F_SML}
  \vspace{-15pt}
\end{figure*}

In this section, we present numerical results based on the system model parameters listed in Table \ref{T_SML}. Other parameters can be chosen arbitrarily so long as the logic outlined throughout the paper remains valid. To evaluate the efficiency of ORIENT, we conduct a comparative analysis with OPT, D3QL, and RND. OPT represents the optimal solution for PIRA, obtained through the use of CPLEX 12.10. D3QL, on the other hand, bears similarities to the method outlined in Algorithm \ref{ORIENT}, but employs exclusively simple linear layers in its DNN, without utilizing any GNNs. This approach forms the foundation for several related studies, including A-DDPG \cite{he_leveraging_2023} and MDRL-SaDS \cite{xuan_multi-agent_2023}, both of which are RL methods designed to enhance the utility of individual requests by considering factors such as resource cost, required bandwidth, and E2E path latency. Lastly, RND represents a random allocation strategy, where resources are allocated to active requests without considering any constraints.

The results are depicted in Fig. \ref{F_SML}, where subfigures A and B represent the average energy consumption per request and total profit across various system sizes, with a constant of $300$ active requests. Here, incrementing the system size entails the creation of a new system graph, incorporating $\mathcal{N} + 1$ network devices and $\mathcal{V} + 1$ compute nodes. Particularly, from $10$ to $13$, resources with a significant energy consumption are introduced into the graph. Between $14$ and $17$, resources with a moderate energy consumption are added to the graph, and the remaining resources included from $18$ to $21$ are characterized by low energy consumption. Subfigures C and D display similar quality metrics, but for different numbers of requests, while keeping the system size fixed at $12$ with equal numbers of resources ($4$) from each level of energy consumption. 

Within the subfigures, it is evident that OPT serves as an upper performance bound, while RND serves as the lower bound. Furthermore, when all resources exhibit high energy consumption rates or are fully occupied (with $\mathcal{N} \leq 13$ in A and $\mathcal{R} \geq 200$ in C), RND shows a similar energy consumption pattern to D3QL-based techniques, but its support rate is limited due to the absence of intelligence and feasibility checks. In contrast, ORIENT excels in both scenarios. As demonstrated in A and B, it achieves near-optimal results by prioritizing high-capacity resources with minimal energy consumption, especially when multiple choices are available for each request ($\mathcal{N} \geq 13$). Similarly, regardless of whether all requests can be supported (C and D, with $\mathcal{R} \leq 200$), near-optimal solutions are consistently attained. However, D3QL exhibits less efficiency and stability compared to ORIENT, mainly due to its inferior state decoding capability.

\section{Conclusion}\label{S_CNC}
In this paper, we examined the joint problem of service instance placement and assignment, path selection, and request prioritization, dubbed PIRA, with the objective of maximizing the overall profit of the system (as a function of the number of supported concurrent requests) while minimizing the overall energy consumption over a continuous period of time, taking into account E2E latency and resource capacity constraints. This problem was formulated as a MINLP problem, its complexity was analyzed, and it was demonstrated that it is an NP-hard problem. Subsequently, a technique named ORIENT was introduced to address the problem in a near-optimal manner, utilizing a GNN-empowered D3QL strategy. The effectiveness of the suggested technique was validated through numerical results. As potential future work, our intention is to tackle the problem in the context of dynamic environments characterized by temporal/spatial fluctuations in requests and resources.

\section*{Acknowledgment}
It is partially supported by the European Union’s Horizon 2020 Research and Innovation Program through the aerOS project under Grant No. 101069732; the Business Finland 6Bridge 6Core project under Grant No. 8410/31/2022; the European Union’s HE research and innovation program HORIZON-JUSNS-2022 under the 6GSandbox project (Grant No. 101096328); and the Research Council of Finland 6G Flagship Programme under Grant No. 346208. This research was also conducted at ICTFICIAL Oy. The paper reflects only the authors’ views, and the commission bears no responsibility for any utilization of the information contained herein.


\bibliographystyle{IEEEtran}
\bibliography{bib}

\end{document}